\documentclass[fleqn,twoside]{article}

\usepackage{amssymb}
\usepackage{graphicx}

\usepackage{amsmath}
\numberwithin{equation}{section}

\begin{document}
\hspace*{13cm} {\large US-03-06}
\vspace{3mm}
%\draft

\begin{center}
%\title{
{\Large\bf Radiative Neutrino Masses in a SUSY GUT Model}

\vspace{3mm}
%\author{
{\bf Yoshio Koide}

%\address{
{\it Department of Physics, University of Shizuoka, 
52-1 Yada, Shizuoka 422-8526, Japan\\
E-mail address: koide@u-shizuoka-ken.ac.jp}

\vspace{2mm}
\date{\today}
\end{center}

\vspace{3mm}
%\maketitle
\begin{abstract}
Radiatively-induced neutrino mass matrix is investigated 
within the framework of an SU(5) SUSY GUT model. 
The model has matter fields of three families 
$\overline{5}_{L(+)i}+5_{L(+)i}$ in addition to the ordinary 
matter fields $\overline{5}_{L(-)i}+10_{L(+)i}$ and 
Higgs fields $H_{(+)}+\overline{H}_{(0)}$,
where $(+,0,-)$  denote the transformation properties 
$( \omega^{+1}, \omega^0, \omega^{-1})$ ($\omega^3=+1$)
under a discrete symmetry Z$_3$.
$R$-parity violating terms are given by 
$\overline{5}_{L(+)} \overline{5}_{L(+)} 10_{L(+)}$,
while the Yukawa interactions are given by
$\overline{H}_{(0)} \overline{5}_{L(-)} 10_{L(+)}$,
i.e. the $\overline{5}$-fields in both are different from
each other.
The Z$_3$ symmetry is only broken by the terms 
$\overline{5}_{L(+)i}5_{L(+)i}$ softly, so that the 
$\overline{5}_{L(+)i}\leftrightarrow \overline{5}_{L(-)i}$
mixings appear at $\mu < M_X$.
Of the $R$-parity violating terms 
$\overline{5}_{L(+)} \overline{5}_{L(+)} 10_{L(+)}$, 
only the terms $(e_L\nu_-\nu_Le_)e_R^c$ sizably appear 
at $\mu < M_X$.
\end{abstract}

%\pacs{

%\maketitle

%%%%%%%%%%%%%%%%%%%%%%%%%%%%%%%%%%%%%%%%%%%%%%%%%%%%%%%%%%%%%
%\begin{multicols}{2}

%\narrowtext

\section{Introduction}
\label{Sec1}

The idea of the radiative neutrino mass \cite{Zee} is an antithesis
to the idea of the neutrino seesaw mechanism \cite{seesaw}: in the
latter model, the neutrinos acquire Dirac masses of the same order
as quark and charged lepton masses and the smallness of the observed
neutrino masses is explained by the seesaw mechanism due to large
Majorana masses of the right-handed neutrinos $\nu_R$, while, in the 
former model, there are no right-handed neutrinos, 
so that there are no Dirac mass terms, and small Majorana 
neutrino masses are generated radiatively. 
Currently, the latter idea is influential,
because it is hard to embed the former model into a grand unification
theory (GUT). A supersymmetric (SUSY) model with $R$-parity violation
can provide neutrino masses \cite{R_SUSY}, but the model cannot be 
embedded into GUT, because the $R$-parity violating terms induce proton
decay inevitably \cite{Smirnov}.

Recently, Sato and the author \cite{K-S} have proposed a model 
with $R$-parity violation within the framework of an SU(5) SUSY GUT. 
In the model,
there are no $R$-parity violating terms $\overline{5}_L \overline{5}_L
10_{L}$ ($\overline{5}_L$ and  $10_{L}$ denote $\overline{5}_L$-plet
and 10-plet matter fields in SU(5) SUSY GUT), which are forbidden by
a discrete symmetry $Z_2$. 
At $\mu < M_X$ ($M_X$ is a unification scale of the SU(5) GUT), 
the Z$_2$ symmetry is softly broken, and 
$\overline{H}_d \leftrightarrow \overline{5}_L$ mixing is
induced, so that the $R$-parity violation terms $\overline{5}'_L 
\overline{5}'_L10_{L}$ are effectively induced from the Yukawa 
interactions $\overline{H}_d \overline{5}_L 10_{L}$.
Although the model is very interesting as an $R$-parity violation
mechanism, it is too restricted for neutrino mass matrix
phenomenology, because the coefficients $\lambda$ of 
$\overline{5}'_L \overline{5}'_L 10_L$ are proportional to
the Yukawa coupling constants $Y_d$ of
$\overline{H}'_d \overline{5}'_L 10_L$.

In contrast to the above scenario, 
in the present paper, we propose another model with $R$-parity violation
within the framework of an SU(5) SUSY GUT:
we have quark and lepton fields $\overline{5}_L +10_L$, which
contribute to the Yukawa interactions as $H_u 10_L 10_L$ and
$\overline{H}_d \overline{5}_L 10_L$; we also have additional 
matter fields $\overline{5}'_L + 5'_L$ which contribute to 
the $R$-parity violating terms $\overline{5}'_L \overline{5}'_L 10_L$.
Since the two $\overline{5}_L$ and $\overline{5}'_L$ are different
from each other, the $R$-parity violating interactions are usually
invisible.
The $R$-parity violating effects become visible only through 
$\overline{5}_L \leftrightarrow \overline{5}'_L$ mixings
in low energy phenomena.

In order to make such a scenario, i.e. in order to allow
the interactions $\overline{5}'_L \overline{5}'_L 10_L$,
but to forbid $\overline{5}_L \overline{5}_L 10_L$ and
$\overline{5}_L \overline{5}'_L 10_L$,
we introduce a discrete symmetry Z$_3$.
(We cannot build such a model by using Z$_2$ symmetry.)
We denote fields with the transformation properties
$\Psi \rightarrow \omega^{+1} \Psi$,
$\Psi \rightarrow \omega^{0} \Psi$ and
$\Psi \rightarrow \omega^{-1} \Psi$ ($\omega^3=+1$) as
$\Psi_{(+)}$, $\Psi_{(0)}$ and $\Psi_{(-)}$, respectively.
We consider matter fields $\overline{5}_{L(-)i} + 10_{L(+)i}$
($i=1,2,3$: family indices) 
which contribute the Yukawa interactions as
\begin{equation}
W_{Y}  =  (Y_u)_{ij} H_{(+)} 10_{L(+)i} 10_{L(+)j} 
+ (Y_d)_{ij} \overline{H}_{(0)}
\overline{5}_{L(-)i}10_{L(+)j} \ ,
%\eqno(1.1)
\end{equation}
and additional matter fields $\overline{5}_{L(+)i} + 5_{L(+)i}$
which contribute the $R$-parity interactions as
\begin{equation}
W_{\not\!R} = \lambda_{ijk} \overline{5}_{L(+)i}
\overline{5}_{L(+)j} 10_{L((+)k} \ .
%\eqno(1.2)
\end{equation}
The $R$-parity violating interactions
$\overline{5}_{L(-)} \overline{5}_{L(-)} 10_{L((+)}$ and 
$\overline{5}_{L(-)} \overline{5}_{L(+)} 10_{L((+)}$ are
forbidden by the Z$_3$ symmetry. 

In order to give $\overline{5}_{L(-)}\leftrightarrow\overline{5}_{L(+)}$
mixings,
\begin{eqnarray}
\overline{5}_{L(-)i} &=& c_i \overline{5}_{Li} +  
s_i \overline{5}'_{Li}  \ ,
\nonumber \\
\overline{5}_{L(+)i} &=& -s_i \overline{5}_{Li}  +  
c_i \overline{5}'_{Li}  \ ,
%\eqno(1.3)
\end{eqnarray}
where $s_i=\sin\theta_i$ and $c_i=\cos\theta_i$, 
we consider a superpotential
\begin{equation}
W_{5} =  \left[ \overline{5}_{L(-)i}(M_5 -g_5 \Phi_{(0)})
 + M_i^{SB} \overline{5}_{L(+)i} \right] 5_{L(+)i}  \ ,
%\eqno(1.4)
\end{equation}
where $\Phi_{(0)}$ is a 24-plet Higgs field with the vacuum
expectation value (VEV) $\langle \Phi_{(0)} \rangle= v_{24}
{\rm diag}(2,2,2,-3,-3)$,
which gives  doublet-triplet splitting
in the mass terms $\overline{5}_{L(-)i} 5_{L(+)i}$ at $\mu < M_X$, 
i.e.
\begin{equation}
M^{(2)} = M_5 +3 g_5 v_{24} \ , \ \ \ 
M^{(3)} = M_5 -2 g_5 v_{24} \ .
%\eqno(1.5)
\end{equation}
The discrete symmetry Z$_3$ is softly broken by 
the $M_i^{SB}$-terms in (1.4).
Then, we obtain
\begin{equation}
W_5 = \sum_{a=2,3} \sqrt{ (M^{(a)})^2 +(M_i^{SB})^2}\,
\overline{5}_{Li} ^{\prime (a)} 5_{L(+)i}^{(a)}  \ ,
%\eqno(1.6)
\end{equation}
where the index $(a)$ denotes that the field $\Psi^{(a)}$
with $a=2$ ($a=3$) is a doublet (triplet) component 
of SU(5)$\rightarrow$SU(2)$\times$SU(3), and
\begin{equation}
s_i^{(a)} = \frac{ M^{(a)} }{ \sqrt{ (M^{(a)})^2 +
(M_i^{SB})^2} } \ ,
\ \ \ 
c_i^{(a)} = \frac{ M_i^{SB} }{ \sqrt{ (M^{(a)})^2 
+(M_i^{SB})^2} } \ .
%\eqno(1.7)
\end{equation}
The field $\overline{5}_{Li}^{\prime (a)}$ has a mass
$\sqrt{ (M^{(a)})^2 +(M_i^{SB})^2}$, while 
$\overline{5}_{Li}^{(a)}$ are massless.
We regard $\overline{5}_{Li} + 10_{L(+)i}$ as the observed 
quarks and leptons at low energy scale ($\mu <M_X$).
Then, the effective $R$-parity violating terms at $\mu< M_X$ are
given by
\begin{equation}
W_{\not\!R}^{eff} = s_i^{(a)} s_j^{(b)} \lambda_{ijk} 
\overline{5}^{(a)}_{Li}\overline{5}^{(b)}_{Lj} 10_{L(+)k} \ .
%\eqno(1.8)
\end{equation}
In order to suppress the unwelcome term $d_R^c d_R^c u_R^c$ 
in the effective $R$-parity violating terms (1.8),
we assume a fine tuning
\begin{equation}
M^{(2)} \sim M_X\ , \ \ \ M^{(3)} \sim m_{SUSY} \ ,
\ \ \ M_i^{SB} \sim M_X \times 10^{-1}\ ,
%\eqno(1.9)
\end{equation}
where $m_{SUSY}$ denotes the SUSY breaking scale 
($m_{SUSY} \sim 1$ TeV) and 
$M^{SB}_i$ are difined by (1.4) (i.e. 
the Z$_3$ symmetry breaking terms are given by
$W_{SB} = M^{SB}_i \overline{5}_{L(+)i} 5_{L(+)i}$
with the mass scale $M_1^{SB} \sim M_2^{SB} \sim M_3^{SB}
\sim 10^{15}$ GeV), so that
\begin{equation}
s_i^{(2)} \simeq 1\ , \ \ 
c_i^{(2)} \simeq \frac{ M_i^{SB} }{ M^{(2)} }\sim  10^{-1}
\ ; \ \ \ \ 
s_i^{(3)} \simeq \frac{ M^{(3)} }{M_i^{SB} } \sim 10^{-12} 
\ , \ \ 
c_i^{(3)} \simeq 1   \ . 
%\eqno(1.10)
\end{equation}
Therefore, the $R$-parity violating terms  $d_R^c d_R^c u_R^c$ and 
$d_R^c (e_L u_L -\nu_L d_L)$ are suppressed by
$s^{(3)}s^{(3)}\sim 10^{-24}$ and $s^{(3)}s^{(2)}\sim 10^{-12}$,
respectively.
Thus, proton decay caused by terms  $d_R^c d_R^c u_R^c$ 
and $d_R^c (e_L u_L -\nu_L d_L)$ is suppressed by a factor
$(s^{(3)})^3s^{(2)}\sim 10^{-36}$.
On the other hand, radiative neutrino masses are generated
by the $R$-parity violating term $(e_L \nu_L -\nu_L e_L)e_R^c$
with a factor $s^{(2)}s^{(2)}\simeq 1$.
The numerical choice (1.9) gives
\begin{eqnarray}
m(\overline{5}_{Li}^{\prime (2)})& \simeq & M^{(2)} \sim M_X \ ,
\nonumber \\
m(\overline{5}_{Li}^{\prime (3)})& \simeq & M^{SB}_i \sim 
M_X \times 10^{-1} \ . 
%\eqno(1.11)
\end{eqnarray}
Since $m(\overline{5}_{Li}^{\prime (3)})< M_X$, the triplet
fields $\overline{5}_{Li}^{\prime (3)}$ can basically contribute 
to the renormalization group equation (RGE) effects at $\mu<M_X$.
However, since we consider $M_i^{SB} \sim M_X\times 10^{-1}$, 
the numerical effect does almost not spoil the gauge-coupling-constant
unification at $\mu=M_X \sim 10^{16}$ GeV.

The up-quark masses are generated by the Yukawa interactions (1.1),
so that we obtain the up-quark mass matrix $M_u$ 
\begin{equation}
(M_u)_{ij} = (Y_u)_{ij} v_u \ ,
%\eqno(1.12)
\end{equation}
where $v_u=\langle H_{(+)}^0\rangle$.
{}From the Yukawa interaction (1.1), we also obtain the
down-quark mass matrix $M_d$ and charged lepton mass
matrix $M_e$ as
\begin{equation}
(M_d^\dagger)_{ij}= c_i^{(3)} (Y_d)_{ij} v_d \ , \ \ \ 
(M_e^*)_{ij} = c_i^{(2)} (Y_d)_{ij} v_d \ ,
%\eqno(1.13)
\end{equation}
i.e.
\begin{equation}
(M_d^\dagger)_{ij}= (c_i^{(3)}/c_i^{(2)})(M_e^*)_{ij}  \ ,
%\eqno(1.14)
\end{equation}
where $v_d=\langle \overline{H}_{(0)}^0\rangle$.
Note that $M_d^T$ has a structure different from $M_e$, 
because the values of $c_i^{(2)}$
can be different from each other.
(The idea  $M_d^T \neq M_e$ based on a mixing between two
$\overline{5}_L$ has been discussed, for example, by
Bando and Kugo \cite{two5mix} in the context of an E$_6$ model.)

In order to give doublet-triplet splitting for the Higgs
fields $H_{(+)}$ and $\overline{H}_{(0)}$, we assume the
``missing partner mechanism" \cite{missing}: for example,
we consider
\begin{equation}
W_H =   \lambda \ H_{(+)} \overline{H}_{50(-)} 
\langle H_{75(0)}\rangle +
\overline{\lambda} \ \overline{H}_{(0)} H_{50(0)}
\langle H_{75(0)}\rangle   \ ,
%\eqno(1.15)
\end{equation}
which gives mass to the triplets in $H_{(+)}+
\overline{H}_{(0)}$, but not to the doublets, where
$H_{50(0)}$ ($\overline{H}_{50(-)}$) and $H_{75(0)}$
are 50-plet and 75-plet Higgs fields, respectively.

%%%%%%%%%%%%%%%%%%%%%%%%%%%%%%%%%%%%%%%%%%%%%%%%%%%%%%%
%\vspace{5mm}
\section{Radiative neutrino mass matrix} 
\label{sec:2}

In this section, we investigate a possible form of the 
radiatively-induced neutrino mass matrix $M_{rad}$.
Contribution from non-zero VEVs 
of sneutrinos $\langle \widetilde{\nu} \rangle \neq 0$ to the 
neutrino mass matrix will be discussed in the next section.

In the present model, since we do not have a term which induces
$\widehat{e}^+_R \leftrightarrow \overline{H}^+_{(0)}$ mixing, 
there is no Zee-type diagram \cite{Zee}, 
which is proportional to the Yukawa vertex 
$(Y_d)_{ij}$ and R-parity violating vertex $\lambda_{ijk}$ .
(The $\widehat{e}^+_R \leftrightarrow \overline{H}^+_{(0)}$ mixing
can come from interactions of a type $\overline{H}\, \overline{H}\,
10_{L(+)}$.  However, in the present model, $\overline{5}$-plet
Higgs fields are only on type $\overline{H}_{(0)}$.
Therefore, the combination $\overline{H}_{(0)}\overline{H}_{(0)}
10_{L(+)}$ is forbidden because of the antisymmetric property
of SU(5) 10-plet fields $10_{L(+)}$.
Even after the SU(5) is broken, 
$\overline{H}_{(0)}^{(2)}\overline{H}_{(0)}^{(2)}$ cannot
couple to the SU(2) singlet $\widehat{e}^+_R$ because
SU(2) singlet composed of $2\times 2$ must be antisymmetric.
Therefore, we cannot bring the $\overline{H}_{(0)}^{(2)}
\overline{H}_{(0)}^{(2)}\widehat{e}^+_R$ term even as a
soft supersymmetry breaking term.)

%%%%%%%%%%%%%%%%%%%%%%%%%%%%%%%%%%%%%%%%%%%%%%%%%%%%%%%%%%%%%%%%%%%%%%
%%%%% penguin %%%%%%%%%%%%%%%%%%%%%%%%%%%%%%%%%%%%%%%%%%%%%%%%%%%%%%%%
%%%%%%%%%%%%%%%%%%%%%%%%%%%%%%%%%%%%%%%%%%%%%%%%%%%%%%%%%%%%%%%%%%%%%%
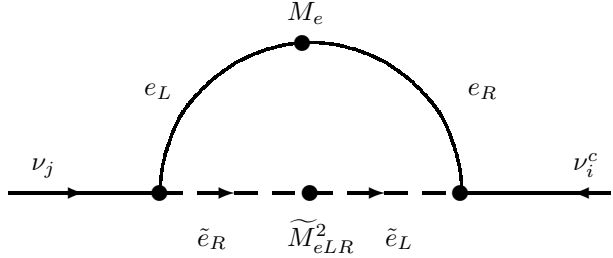
\begin{figure}[hb]
\unitlength=1cm
%\begin{picture}(16.6,4)
\begin{center}
\begin{picture}(9,4)
\thicklines
%
% lepton type
%
% nu_i
\put(0.5,1){\line(1,0){2}}
\put(1.5,1){\vector(1,0){0}}
\put(0.8,1.3){$\nu_j$}
% vertex
\put(2.5,1){\circle*{0.2}}
% slepton
\multiput(2.5,1)(0.5,0){8}{\line(1,0){0.3}}
\put(3.5,1){\vector(1,0){0}}
\put(3,0.3){$\tilde{e}_R$}
% vertex
\put(4.5,1){\circle*{0.2}}
% slepton
\put(5.5,1){\vector(1,0){0}}
\put(5.5,0.3){$\tilde{e}_L$}
% vertex
\put(6.5,1){\circle*{0.2}}
\put(4.2,0.3){$\widetilde{M}^{2}_{eLR}$}
% neutrino
\put(6.5,1){\line(1,0){2}}
\put(8,1){\vector(-1,0){0}}
\put(8,1.3){$\nu_i^c$}
%%%% charged lepton circle %%%%%
\put(4.5,1){
\qbezier(-2,0)(-2.01,0.35)(-1.88,0.68)
\qbezier(-1.88,0.68)(-1.78,1.03)(-1.53,1.29)
\qbezier(-1.53,1.29)(-1.3,1.55)(-1,1.73)
\qbezier(-1,1.73)(-0.69,1.9)(-0.35,1.97)
\qbezier(-0.35,1.97)(0,2.03)(0.35,1.97)
\qbezier(2,0)(2.01,0.35)(1.88,0.68)
\qbezier(1.88,0.68)(1.78,1.03)(1.53,1.29)
\qbezier(1.53,1.29)(1.3,1.55)(1,1.73)
\qbezier(1,1.73)(0.69,1.9)(0.35,1.97)
}
\put(4.4,3){\circle*{0.2}}
\put(4.2,3.3){$M_e$}
\put(2.3,2.3){$e_L$}
\put(6.6,2.3){$e_R$}
%
% down type
%
% nu_i
%\put(8.8,1){\line(1,0){2}}
%\put(9.8,1){\vector(1,0){0}}
%\put(9.1,1.3){$\nu_j$}
% vertex
%\put(10.8,1){\circle*{0.2}}
% slepton
%\multiput(10.8,1)(0.5,0){8}{\line(1,0){0.3}}
%\put(11.8,1){\vector(1,0){0}}
%\put(11.3,0.3){$\tilde{d}_R$}
% vertex
%\put(12.8,1){\circle*{0.2}}
% slepton
%\put(13.8,1){\vector(1,0){0}}
%\put(13.8,0.3){$\tilde{d}_L$}
% vertex
%\put(14.8,1){\circle*{0.2}}
%\put(12.5,0.3){$\widetilde{M}^{2}_{dLR}$}
% neutrino
%\put(14.8,1){\line(1,0){2}}
%\put(16.3,1){\vector(-1,0){0}}
%\put(16.3,1.3){$\nu_i^c$}
%%%% downquark circle %%%%%
%\put(12.8,1){
%\qbezier(-2,0)(-2.01,0.35)(-1.88,0.68)
%\qbezier(-1.88,0.68)(-1.78,1.03)(-1.53,1.29)
%\qbezier(-1.53,1.29)(-1.3,1.55)(-1,1.73)
%\qbezier(-1,1.73)(-0.69,1.9)(-0.35,1.97)
%
%\qbezier(-0.35,1.97)(0,2.03)(0.35,1.97)
%
%\qbezier(2,0)(2.01,0.35)(1.88,0.68)
%\qbezier(1.88,0.68)(1.78,1.03)(1.53,1.29)
%\qbezier(1.53,1.29)(1.3,1.55)(1,1.73)
%\qbezier(1,1.73)(0.69,1.9)(0.35,1.97)
%}
%\put(12.7,3){\circle*{0.2}}
%\put(12.5,3.3){$M_d$}
%\put(10.6,2.3){$d_L$}
%\put(14.9,2.3){$d_R$}
\end{picture}
\caption{Radiative generation of neutrino Majorana mass}
\label{fig:numass}
\end{center}
\end{figure}
%%%%%

Only the radiative neutrino masses in the present scenario come
from a charged-lepton loop diagram:
the radiative diagram with 
$(\nu_L)_j \rightarrow (e_R)_l + (\widetilde{e}_L^c)_n$ and
$(e_L)_k + (\widetilde{e}_L^c)_m \rightarrow (\nu_L^c)_i$.
The contributions $(M_{rad})_{ij}$ from the charged lepton 
loop are given,
except for the common factors, as follows:
%%%%%%%%%%%%%%%%%%%%%%%%%%%%%%%%%%%%%%%%%%%%%%%%%%%%%%%%%%%%%%%%%%
\begin{equation}
(M_{rad})_{ij}= 
s_i s_j s_k s_n \lambda_{ikm} \lambda_{jnl}
  (M_e)_{kl} (\widetilde{M}_{eLR}^{2T})_{mn} + 
(i \leftrightarrow j)  \ ,
%\eqno(2.1)
\end{equation}
where $s_i=s_i^{(2)}$, $m_i=m(e_i)=(m_e , m_\mu , m_\tau)$ and
$M_e$ and $\widetilde{M}_{eLR}^2$ are charged-lepton and 
charged-slepton-LR mass matrices, respectively.
Since  $\widetilde{M}_{eLR}^2$ is proportional to $M_e$, i.e., 
$\widetilde{M}^2_{eLR} = A  M_e$
($A$ is the coefficient of the soft SUSY breaking terms
$(Y_d)_{ij} (\tilde{\nu}, \tilde{e})^T_{Li} \tilde{e}_{Lj}^c
\overline{H}_{(0)}$ with $A \sim 1$ TeV), we obtain
\begin{equation}
(M_{rad})_{ij} = 2A
s_i s_j s_k s_n \lambda_{ikm} \lambda_{jnl} (M_e)_{kl} (M_e)_{nm}
 \ .
%\eqno(2.2)
\end{equation}
Therefore, the mass matrix $M_{rad}$ on the basis 
with $M_e=D_e \equiv {\rm diag}(m_e, m_\mu, m_\tau)$ is given by
\begin{equation}
(M_{rad})_{ij} = m_0^{rad}
s_i s_j s_k s_l \lambda_{ikl} \lambda_{jlk} 
\frac{m_k m_l}{m_3^2}  \ ,
%\eqno(2.3)
\end{equation}
where
\begin{equation}
m_0^{rad} = \frac{2}{16 \pi^2} A m_3^2
 F(m_{\tilde{e}_R}^2, m_{\tilde{e}_L}^2)  \ ,
%\eqno(2.4)
\end{equation}
\begin{equation}
F(m_a^2, m_b^2)= \frac{1}{m_a^2-m_b^2} \ln\frac{m_a^2}{m_b^2} \ .
%\eqno(2.5)
\end{equation}

Since the coefficient $\lambda_{ijk}$ is antisymmetric in the permutation
$i \leftrightarrow j$, it is useful to define
\begin{equation}
\lambda_{ijk} = \varepsilon_{ijl} h_{lk} \ ,
%\eqno(2.6)
\end{equation}
and
\begin{equation}
H_{ij} =  h_{ij} m_j s_j \ .
%\eqno(2.7)
\end{equation}
Then, we can rewrite (2.4) as 
\begin{equation}
(M_{rad})_{ij} = \frac{m_0^{rad}}{m^2_3} s_i s_j \varepsilon_{ikm} 
\varepsilon_{jln} H_{ml} H_{nk} \ .
%\eqno(2.8)
\end{equation}
The expression (2.8) is explicitly given as follows:
\begin{equation}
M_{11} = s^2_1 \left[ H^2_{23} + H^2_{32} -2H_{22} H_{33} \right] \ ,
%\eqno(2.9)
\end{equation}
\begin{equation}
M_{22} = s^2_2 \left[ H^2_{31} + H^2_{13} -2H_{33} H_{11} \right] \ ,
%\eqno(2.10)
\end{equation}
\begin{equation}
M_{33} = s^2_3 \left[ H^2_{12} + H^2_{21} -2H_{11} H_{22} \right] \ ,
%\eqno(2.11)
\end{equation}
\begin{equation}
M_{12} = M_{21} = s_1 s_2
\left[ (H_{12} + H_{21})H_{33} -H_{23}H_{13} -H_{32} H_{31} \right] \ ,
%\eqno(2.12)
\end{equation}
\begin{equation}
M_{13} = M_{31} = s_1 s_3
\left[ (H_{13} + H_{31})H_{22} -H_{23}H_{21} -H_{32} H_{12} \right] \ ,
%\eqno(2.13)
\end{equation}
\begin{equation}
M_{23} = M_{32} = s_2 s_3
\left[ (H_{23} + H_{32})H_{11} -H_{31}H_{21} -H_{13} H_{12} \right] \ ,
%\eqno(2.14)
\end{equation}
where $M_{ij} \equiv (M_{rad})_{ij}$ and we have dropped a common factor
$m_0^{rad} / m^2_3$. As discussed in (1.10), 
in a phenomenological investigation
in the next section,  we will take $s_1 = s_2 = s_3 = 1$ for simplicity.

%%%%%%%%%%%%%%%%%%%%%%%%%%%%%%%%%%%%%%%%%%%%%%%%%%%%%%%
%\vspace{5mm}
\section{Phenomenology} 
\label{sec:3}

In general, the sneutrinos $\widetilde{\nu}_i$ can have VEVs 
$v_i \equiv \langle \widetilde{\nu}_i \rangle \neq 0$
\cite{snu}. Since the
mass matrix for $(\nu_1, \nu_2 , \nu_3 , \widetilde{W}^0)$ (except
for the radiative masses) is given by
\begin{equation}
\left(
\begin{array}{cccc}
0 & 0 & 0 & \frac{1}{2}g v_1 \\
0 & 0 & 0 & \frac{1}{2}g v_2 \\
0 & 0 & 0 & \frac{1}{2}g v_3 \\
\frac{1}{2}g v_1 & \frac{1}{2}g v_2 
& \frac{1}{2}g v_3 & M_{\widetilde{W}}
\end{array} \right) \ ,
%\eqno(3.1)
\end{equation}
where, for simplicity, we have dropped the elements for 
$\widetilde{B}^0$, 
the contribution $M_{\tilde{\nu}}$ from 
$\langle \widetilde{\nu}_i \rangle \neq 0$
to the neutrino masses is expressed by 
\begin{equation}
M_{\widetilde{\nu}} 
\simeq 
- \frac{g^2}{4}
\left(
\begin{array}{c}
v_1  \\
v_2  \\
v_3 
\end{array} \right) 
(M_{\widetilde{W}})^{-1}
(v_1 \ v_2 \ v_3)
= -\frac{g^2}{4 M_{\widetilde{W}}}
\left(
\begin{array}{ccc}
v^2_1 & v_1 v_2 & v_1 v_3 \\
v_1 v_2 & v^2_2 & v_2 v_3 \\
v_1 v_3 & v_2 v_3 & v_3^2 
\end{array} \right) 
\ ,
%\eqno(3.2)
\end{equation}
under the seesaw approximation.
Note that the matrix $M_{\widetilde{\nu}}$ is a rank-1 matrix.
Therefore, in the present model, the neutrino mass matrix $M_\nu$
is given by
\begin{equation}
M_{\nu} = M_{rad} + M_{\widetilde{\nu}} \ .
%\eqno(3.3)
\end{equation}

We have estimated the absolute magnitudes of the radiative
masses in (2.3)--(2.5).
On the other hand, it is hard to estimate the absolute values
of $\langle \widetilde{\nu}_i \rangle$, because,
in the present model, there is neither a term corresponding to
the so-called ``$\mu$-term" $\mu \overline{H}_d H_u$ nor
$\overline{5}_{L(-)i} \leftrightarrow \overline{H}_{(0)}$
mixing terms, so that the sneutrinos $\widetilde{\nu}_i$ cannot
have the VEVs $\langle \widetilde{\nu}_i \rangle$ at the
tree level.
The non-zero VEVs appears only through the renormalization 
group equation (RGE) effect\cite{evolvev}.
The contribution highly depends on an explicit model of
the SUSY breaking.
Therefore, in the present paper, we will deal with 
the relative ratio of the contributions $M_{\tilde{\nu}}$ 
to $M_{rad}$ as a free parameter.

The recent neutrino data \cite{solar,kamland,atm}
have indicated that 
$\sin^2 2\theta_{atm} \simeq 1$ and $\tan^2 \theta_{solar} \simeq 0.5$.
In response to these observations, He and Zee have found 
a phenomenological neutrino mass matrix \cite{He-Zee}
\begin{equation}
M_{\nu} = m_0
\left(
\begin{array}{ccc}
2+x & 0 & 0 \\
0 & 1-y+x & 1+y \\
0 & 1+y & 1-y+x 
\end{array} \right) 
+ m_0 \varepsilon
\left(
\begin{array}{ccc}
1 & 1 & 1 \\
1 & 1 & 1 \\
1 & 1 & 1 
\end{array} \right) 
\ ,
%\eqno(3.4)
\end{equation}
which leads to a nearly bimaximal mixing 
\begin{equation}
U =
\left(
\begin{array}{ccc}
\frac{2}{\sqrt6} & \frac{1}{\sqrt3} & 0 \\
-\frac{1}{\sqrt6} & \frac{1}{\sqrt3} & -\frac{1}{\sqrt2} \\
-\frac{1}{\sqrt6} & \frac{1}{\sqrt3} & \frac{1}{\sqrt2} 
\end{array} \right) \ ,
%\eqno(3.5)
\end{equation}
i.e. 
\begin{equation}
\sin^2 2\theta_{atm} = 1 \ ,
%\eqno(3.6)
\end{equation}
\begin{equation}
\tan^2 \theta_{solar} = \frac{1}{2} \ .
%\eqno(3.7)
\end{equation}
(Although He and Zee gave the mass matrix (3.4) with $x=0$
in Ref.~\cite{He-Zee}, since a term
which is proportional to a unit matrix does not affect the mixing matrix
form, the most general form of the He--Zee mass matrix is given by (3.4).)
The mass matrix (3.4) gives the following mass eigenvalues:
\begin{eqnarray}
m_{\nu 1} &=& m_0 (2+x)  \ ,
\nonumber \\
m_{\nu 2} &=& m_0 (2+x+3\varepsilon)  \ ,  \\
m_{\nu 3} &=& m_0 (x - 2y) \ , \nonumber
%\eqno(3.8)
\end{eqnarray}
and
\begin{eqnarray}
\Delta m^2_{21} = m^2_{\nu 2} - m^2_{\nu 1} = 12 \varepsilon 
\left(1+ \frac{1}{2}x + \frac{3}{4} \varepsilon \right) m^2_0 \ ,
%\eqno(3.9)
\end{eqnarray}
\begin{eqnarray}
\Delta m^2_{32} = m^2_{\nu 3} - m^2_{\nu 2} = -4 
\left(1+ x -y + \frac{2}{3} \varepsilon \right) 
\left(1+ y + \frac{3}{2} \varepsilon \right)
m^2_0 \ ,
%\eqno(3.10)
\end{eqnarray}
\begin{equation}
R = \frac{\Delta m^2_{21}}{\Delta m^2_{32}} 
= -
\frac{3(2+x+3\varepsilon/2)\varepsilon}
{2(1+x-y)(1+y)}
 \ .
%\eqno(3.11)
\end{equation}
(Therefore, the parameter $y$ has to be $y \neq -1$ and $y \neq 1+x$.)

In the present model, there are many adjustable parameters for
the neutrino mass matrix phenomenology.
Let us seek for an example with simple and plausible forms of
$M_{rad}$ and $M_{\tilde{\nu}}$ with a clue of the successful
He--Zee mass matrix form.
First, we think that it is likely that the VEVs 
$\langle\tilde{\nu}_i\rangle$ 
are democratic on the basis on
which the charged lepton mass matrix is diagonal, i.e.
\begin{equation}
\langle\tilde{\nu}_1\rangle  = \langle\tilde{\nu}_2\rangle  
= \langle\tilde{\nu}_3\rangle  \ ,
%\eqno(3.12)
\end{equation}
so that we can regard the second term in the He--Zee matrix (3.4)
as $M_{\tilde{\nu}}$ which
originates in the sneutrino VEVs. 
Then, it is interesting
whether our radiative mass matrix (2.8) can give the first term in the
He--Zee mass matrix (3.4) or not. 

Corresponding to the assumption (3.12), we may also suppose that the 
coefficients $h_{ij}$ are invariant under the permutation among 
$\ell_{Li} = (\nu_{Li} \ , e_{Li})$ which belong to $\overline{5}_{Li}$
(not among $e^c_{Ri}$ which belong to $10_{Li}$). The most simple case will
be
\begin{equation}
h = \lambda 
\left(
\begin{array}{ccc}
1 & 0 & 0 \\
1 & 0 & 0 \\
1 & 0 & 0 
\end{array} \right) \ .
%\eqno(3.13)
\end{equation}
Then, we obtain the radiative neutrino mass matrix
\begin{equation}
M_{rad} =m_0^{rad} \lambda^2 \frac{m^2_1}{m^2_3} 
\left(
\begin{array}{ccc}
0 & 0 & 0 \\
0 & 1 & -1 \\
0 & -1 & 1 
\end{array} \right) \ ,
%\eqno(3.14)
\end{equation}
which corresponds the first term in the He--Zee mass matrix 
(3.4) with
$x = -2$ and $y = -2$, so that we get
\begin{equation}
R \simeq \frac{9}{4} \varepsilon^2 \ ,
%\eqno(3.15)
\end{equation}
where 
\begin{equation}
\varepsilon = -\frac{g^2}{4} 
\frac{\langle \tilde{\nu} \rangle^2}{M_{\tilde{W}} m_0} \ ,
%\eqno(3.16)
\end{equation}
\begin{equation}
m_0 = m_0^{rad} \lambda^2 \frac{m_e^2}{m_\tau^2} \ .
%\eqno(3.17)
\end{equation}
{}From the best fit values of $\Delta m^2_{ij}$ 
\cite{solar,kamland,atm},
\begin{equation}
R_{obs} = \frac{6.9 \times 10^{-5} \ {\rm eV^2}}
{2.5 \times 10^{-3} \ {\rm eV^2}}
= 2.76 \times 10^{-2} \ ,
%\eqno(3.18)
\end{equation}
we obtain
\begin{equation}
\varepsilon = 0.111 \ ,
%\eqno(3.19)
\end{equation}
and
\begin{equation}
m_{\nu 1} = 0 \ , \ \ \  m_{\nu 2} = 0.0083 \ {\rm eV} \ ,
\ \ \ m_{\nu 3} = 0.050 \ {\rm eV} \ , \ \ \ 
(m_0 =0.025 \ {\rm eV} \ ,
%\eqno(3.20)
\end{equation}
where we have used the best fit values \cite{solar,kamland,atm}
$\Delta m^2_{atm}=2.5 \times 10^{-3}$ eV$^2$ and 
$\Delta m^2_{solar}=6.9 \times 10^{-5}$ eV$^2$.
In the present model (3.14), the absolute magnitude of $m_{\nu 3}$,
which is radiatively generated, is given by
\begin{equation}
m_{\nu 3} = \frac{1}{4\pi} AF \lambda^2 m_e^2 
= 1.9\times 10^{-2} \hat{A} \hat{F} \lambda^2 \ {\rm eV}\ ,
%\eqno(3.21)
\end{equation}
where $\hat{A}$ and $\hat{F}$ are numerical values of 
the parameters $A$ and $F$ in unit of TeV, which are defined by 
$\widetilde{M}^2_{eLR} =A M_e$ and the equation (2.5), 
respectively.
If we, for example, take $A\simeq 1$ TeV and $\tilde{m}^2_{eL}
\simeq \tilde{m}^2_{eR} \simeq 0.5$ TeV, we obtain
$m_{\nu 3} \simeq 0.075 \lambda^2$ eV.
Thus, roughly speaking, the choice $m_{SUSY} \sim 1$ TeV and 
$\lambda\sim 1$ can give a reasonable magnitude of $m_{\nu 3}$.

%%%%%%%%%%%%%%%%%%%%%%%%%%%%%%%%%%%%%%%%%%%%%%%%%%%%%%%%%%%%%%%%%

%%%%%%%%%%%%%%%%%%%%%%%%%%%%%%%%%%%%%%%%%%%%%%%%%
%\vspace{5mm}
\section{Summary} 
\label{sec:5}

 In conclusion, we have proposed a model with 
$R$-parity violation within
the framework of an SU(5) SUSY GUT. In the model, we have matter fields
$10_{L(+)} + \overline{5}_{L(-)} + 5_{L(+)} + \overline{5}_{L(+)}$ and
Higgs fields $H_{(+)}$ and $\overline{H}_{(0)}$ , 
where $(+,\ 0, \ -)$ denote
their transformation properties 
$(\omega^{+1}, \ \omega^{0}, \ \omega^{-1})$
under a discrete symmetry Z$_3$, respectively. 
Although $\overline{5}_{L(-)}  5_{L(+)}$ 
acquires a heavy mass $M_5$ at $\mu = M_X$, the effective
masses of the triplet and doublet components $\overline{5}^{(3)}_{L(-)}  
5^{(3)}_{L(+)}$ and $\overline{5}^{(2)}_{L(-)}  \overline{5}^{(2)}_{L(+)}$,
$M^{(3)}$ and $M^{(2)}$, are given by $M^{(3)} \sim M_W$ and $M^{(2)} \sim 
M_X$, respectively, because we consider a fine tuning term $g_5 
\overline{5}_{L(-)} \Phi_{(0)} 5_{L(+)}$ with VEVs $\langle \Phi_{(0)} 
\rangle = v_{24}(2, 2, 2, -3, -3)$. 
At an intermediate energy scale
$\mu = M_I \sim 10^{15}$ GeV, the Z$_3$ symmetry is broken by
the term $M^{SB}\overline{5}_{L(+)}  5_{L(+)}$, 
so that masses of 
$\overline{5}^{(3)}_{L(-)}$ and  $\overline{5}^{(2)}_{L(-)}$ are given by 
$m(\overline{5}^{(3)}_{L(+)}) \simeq M^{SB} \sim M_I$ and 
$m(\overline{5}^{(2)}_{L(-)}) \simeq M^{(2)} \sim M_X$. 
In other words, at a low 
energy scale, the massless matter fields are 
$\overline{5}^{(3)}_{L(-)}  +
\overline{5}^{(2)}_{L(+)} + 10_{L(+)}$. 
Therefore,  the $R$-parity violating 
interactions $\overline{5}_{L(+)} \overline{5}_{L(+)} 10_{L(+)}$ 
are invisible 
in the triplet sector, while those are visible in the doublet sector. 
Since 
we take the fine tuning parameters $M^{(3)}$,  $M^{(2)}$ and $M^{SB}$ as 
$M^{(3)} \sim m_{SUSY}$ , $M^{(2)} \sim M_X$ and $M^{SB} \sim M_X \times 
10^{-1}$, the mixing angles $\theta^{(a)}_i$ between 
$\overline{5}^{(a)}_{L(+)i}$ and $\overline{5}^{(a)}_{L(-)i}$ 
(the observed 
quarks and leptons $\overline{5}_L$ are defined as 
$\overline{5}_{Li} = c_i
\overline{5}_{L(-)i} - s_i \overline{5}_{L(+)i}$) are given by 
$s^{(3)}_{i}
\simeq M^{(3)} / M^{SB}_i \sim 10^{-12}$ and $c^{(2)}_i \simeq M^{SB}_i / 
M^{(2)} \sim 10^{-1}$, i.e. the triplet components 
in the effective $R$-parity
violating interactions $\overline{5}_L \overline{5}_L 10_{L(+)}$ are highly
suppressed by the factors $s_i^{(3)} \sim 10^{-12}$, while the doublet 
components are visible because of $s^{(2)}_i \sim 1$.

In the present model, the radiative neutrino masses are generated 
only through the charged lepton loop. 
The general radiative mass matrix form $M_{rad}$ is given
by the expression (2.8) [(2.9)--(2.14)]. 
If there are contributions 
$M_{\widetilde{\nu}}$ from  VEVs of the sneutrinos $\langle \widetilde{\nu} 
\rangle \neq 0$ to the neutrino mass matrix $M_\nu$ with suitable magnitudes
relative to $M_{rad}$, especially, with a democratic form (3.12), we can obtain
the He-Zee neutrino mass matrix form (3.4), which leads to a nearly bimaximal
mixing with $\sin^2 2\theta_{atm} = 1$ and $\tan^2 \theta_{solar} = 1/2$. 
Of course, this is only an example of the explicit mass matrix form and
the He--Zee matrix with forms (3.12) and (3.14) are not a logical consequence 
of the present model. 
We have to assume something of an anzatz for a flavor symmetry.
Maybe, a more plausible ansatz for the flavor symmetry will 
give a more elegant mass matrix form which gives beautiful explanations 
for the observed neutrino and lepton-flavor-violation phenomena. 
Search for such a flavor symmetry is one of our future tasks.

In the present paper, we did not discuss the quark and charged lepton mass 
matrices. 
In the present model, the down-quark mass matrix $M_d$ is related to
the charged lepton mass matrix $M_e$ as $M^T_d = C M_e$ with $C \neq {\bf 1}$.
Investigation of a possible structure of $C$ is also 
a future task in the model.

It is interesting to extend the model to a further large unification group. 
In 
the present SU(5) model, we have two types of the matter fields with the 
transformation properties $\omega^{+1}$ and $\omega^{-1}$ 
under the discrete 
symmetry Z$_3$, i.e. $\overline{5}_{L(+)} + 10_{L(+)}$ 
and $\overline{5}_{L(-)}  + 5_{L(+)}$. 
For example, if we suppose an SU(10) model, we can regard 
$\overline{5}_{L(+)} + 10_{L(+)}$ [$+1_{L(+)}$] and 
$\overline{5}_{L(\pm)} + 5_{L(\pm)}$ as $16_{(+)}$ 
and $10_{(\pm)}$ of SO(10), respectively. 
We are also interested in a 27-plet representation of E$_6$ , 
which is decomposed into $16+10+1$ of SO(10). 
Thus,  the present model has a possibility of a further 
extension.

In conclusion, the present model will bring fruitful results not only in 
phenomenology, but also in a theoretical extension.

%%%%%%%%%%%%%%%%%%%%%%%%%%%%%%%%%%%%%%%%%%%%%%%%
\vspace{7mm}

\centerline{\large\bf Acknowledgments}
The present work was an improved version of the paper
hep-ph/0210188 (unpublished) in collaboration
with J.~Sato, and completed with a valuable hint
from the recent paper hep-ph/0305291 \cite{K-S} 
in collaboration with J.~Sato.
The author would like to thank J.~Sato for helpful
conversations.
He also thank M.~Yamaguchi and M.~Bando for informing
him useful references.
This work was supported by the Grant-in-Aid for
Scientific Research, the Ministry of Education,
Science and Culture, Japan (Grant Number 15540283).

%%%%%%%%%%%%%%%%%%%%%%%%%%%%%%%%%%%%%%%%%%%%

%\end{multicols}

\end{document}